\documentclass[preprint,showpacs,preprintnumbers,amsmath,amssymb]{revtex4}
\usepackage{graphicx}
\usepackage{dcolumn}
\usepackage{bm}
\usepackage{amssymb}
\usepackage{amsmath}
\usepackage{latexsym}

\begin{document}

\title{Application of Regge-theory to astronomical objects}

\author{G.G. Adamian$^{1}$,  N.V. Antonenko$^{1}$,   H. Lenske$^{2}$, and V.V. Sargsyan$^{1}$}
\affiliation{
$^{1}$Joint Institute for Nuclear Research, 141980 Dubna, Russia,\\
$^{2}$Institut f\"ur Theoretische Physik der
Justus--Liebig--Universit\"at, D--35392 Giessen, Germany}

\date{\today}

\begin{abstract}
Using the model based on the Regge-like laws,
new analytical formulas  are obtained for the moment of inertia, the rotation frequency, and the radius of
astronomical non-exotic objects (planets, stars, galaxies, and clusters of galaxies).
 The rotation frequency and moment of inertia of   neutron star
and the observable Universe are estimated. The estimates of the average numbers of stars and galaxies in the observable Universe are given.
The Darwin instability effect in the binary systems  (di-planets, di-stars, and di-galaxies)
is also analyzed.

\end{abstract}


\maketitle

\section{Introduction}
\label{sec:Intro}

The Regge-theory proved to be very influential in the development of elementary-particle physics \cite{Regge,Regge2,CFplot}.
As  known, one of the most important properties of elementary particles is their ability to have an innate proper spin.
The observed correlation between the spin and mass of hadrons shows that the heavier the hadron, the greater the spin it can have.
The relation between the maximum spin $S$ and mass $M$ for all known hadrons and hadron resonances is given by a rectilinear Regge trajectory
in the doubly logarithmic plane $\log_{10}(M)$--$\log_{10}(S)$ (the Chew-Frautschi plot), which for large spin values can be represented as
\begin{eqnarray}
S=\hbar\left(\frac{M}{m_p}\right)^{2},
\label{eq_0}
\end{eqnarray}
where  $\hbar$ and $m_p$   are the  Planck constant and mass of  proton, respectively \cite{CFplot}.
In Refs. \cite{Murad1,Murad2,Murad3}, the cosmic analog of the Chew-Frautschi plot with two important cosmological
Eddington and Chandrasekhar points on it has been constructed.
As shown in Refs. \cite{Murad1,Murad2,Murad3}, in the general case of   $n$--dimensional astrophysical object
the relation between
its spin $S$ and mass $M$ is the following:
\begin{eqnarray}
S=\hbar\left(\frac{M}{m_p}\right)^{1+1/n}.
\label{eq_1}
\end{eqnarray}
 The quasi-classical formula (\ref{eq_1}) has been derived from simple dimensionality considerations
 and the requirement of similarity with formula (\ref{eq_0}) \cite{Murad1,Murad2,Murad3}.
  As seen, for the one-dimensional case $n=1$, Eq. (\ref{eq_1}) turns into Eq. (\ref{eq_0}).
 In contrast to earlier semi-phenomenological
 approaches these expressions contain only fundamental constants
as the parameters and are independent of any fitted  empirical quantities.
The application of Regge ideas to astrophysics  has   shown that the spins $S$ of   planets and stars are well described
by the Regge-like law for a sphere ($S\sim M^{4/3}$, $n=3$), while the spins of galaxies
and clusters of galaxies obey the Regge-like law for a disk ($S\sim M^{3/2}$, $n=2$) \cite{Murad1,Murad2,Murad3}.
The proposed simple Regge-like law  allows us to obtain reasonable numerical values
for  spins of cosmic objects in a self-consistent manner, starting from planets and ending with the astronomical
Universe as a whole, in an extremely wide range of masses (30 orders of magnitude) and spins (50 orders of magnitude)  \cite{Murad1,Murad2,Murad3}.
In addition, Refs. \cite{Murad1,Murad2,Murad3} offer  an explanation of the origin of cosmic  objects and their rotation
in the framework of the concept of Regge trajectories and the Ambartsumian cosmogony \cite{Ambart,Ambart1,Ambart2}.

The aim of the present article is to obtain the analytical expressions for the moment of inertia, the rotation frequency, and the radius of celestial
objects and to study the Darwin instability effect in the binary star or  binary galaxy \cite{Darwin:1879} by using
the well-known model of Refs. \cite{Murad1,Murad2,Murad3} based on the concept  of Regge trajectories.


\section{Planets, stars, galaxies, and clusters of galaxies}
\label{sec:Theory1}

Let us consider the individual astronomical objects  (planets, stars, galaxies, and clusters of galaxies).
The general virial theorem  is
\begin{eqnarray}
U=-2(E_k+V_r),
\label{eq_2}
\end{eqnarray}
where $U=-\frac{\omega GM^2}{R}$ is the gravitational potential with the gravitational constant $G$,
 the radius  $R$,  and the dimensionless structural factor $\omega$ of object. The value of $E_k$ includes the
kinetic energy of both the thermal motion of particles and
the macroscopic motions of matter (caused by pulsations, convective currents, etc.),
with the exception of the rotation energy $V_r=\frac{S^2}{2\Im}$ of the object.
The dimensionless structural factor
\begin{eqnarray}
\omega=\int_0^1dq_x\frac{q_x}{x}
\label{eq_21}
\end{eqnarray}
is determined by the density profile $\rho(r)$ of the object.
Here, the $x=r/R$ and
$$
q_x=M_x/M=\int_0^rdr'r'^2\rho(r')/\int_0^Rdr'r'^2\rho(r')
$$
are   the fraction of the radius and the mass fraction, respectively,   of the object at a distance $r$ from the center of object.
Note that because of $q_x\le 1$, we have $\omega\ge 1/2$. For the homogeneous density distribution, $\omega=3/5$.
If the concentration of matter indefinitely increases to the center of object, then  $\omega=99/125\approx 0.8$.
The structure of main sequence stars are rather well distributed by a polytropes of indexes $n_0$ from 1.5 to about 3.5
and corresponding structural factors $\omega=3/(5 - n_0)$ from 6/7 to about 2.

Using Eqs. (\ref{eq_1}), (\ref{eq_2}),
and the observed scaling law
\begin{eqnarray}
R=\alpha M^m
\label{eq_3}
\end{eqnarray}
between the radius $R$ and mass $M$ of object ($\alpha$ and $m$  are constants),
we obtain the expression for the moment of inertia
\begin{eqnarray}
\Im=-\frac{S^2}{U+2E_k}=\frac{\hbar^2\alpha}{\omega GM^{2-m}-2\alpha E_k}\left(\frac{M}{m_p}\right)^{2+2/n}.
\label{eq_4}
\end{eqnarray}
%
%
%
On other side,
\begin{eqnarray}
\Im=\frac{S}{\Omega}=\frac{\hbar}{\Omega}\left(\frac{M}{m_p}\right)^{1+1/n}.
\label{eq_4-2}
\end{eqnarray}
Assuming that the rotational frequency $\Omega$
is a function  of mass $M$ ($\gamma$ and $q$ are constants),
\begin{eqnarray}
\Omega=\gamma M^q,
\label{eq_4-2-O}
\end{eqnarray}
 and using formulas  (\ref{eq_4}) and (\ref{eq_4-2}),
we derive $q=0$
and the following relation
\begin{eqnarray}
m=1-\frac{1}{n},
\label{eq_6}
\end{eqnarray}
connecting the  constants $m$ and $n$, or  the radius-mass relation
\begin{eqnarray}
R=\alpha M^{1-1/n},
\label{eq_31}
\end{eqnarray}
and new formulas
\begin{eqnarray}
E_k=\beta\left(\frac{M}{m_p}\right)^{1+1/n},
\label{eq_41-1}
\end{eqnarray}
\begin{eqnarray}
\Im=\frac{\hbar^2\alpha}{\omega Gm_p^{1+1/n}-2\alpha\beta}\left(\frac{M}{m_p}\right)^{1+1/n},
\label{eq_41}
\end{eqnarray}
\begin{eqnarray}
\Omega=\frac{\omega Gm_p^{1+1/n}}{\hbar\alpha}-\frac{2\beta}{\hbar}.
\label{eq_51}
\end{eqnarray}
The value of $\beta$ in (\ref{eq_41-1}) is the constant which can be extracted from the observed data of $\Omega$, $\alpha$, and $\omega$.
As seen, the moment of inertia (\ref{eq_41})  depends on the dimension $n$ of the object,
  the classical and quantum fundamental constants $G$, $\hbar$, $m_p$, the dimensionless structural factor $\omega$,
and  the constants   $\alpha$ and $\beta$   determined from the observed data.
The Regge-like spin-mass formula (\ref{eq_1}), the moment-of-inertia-mass formula (\ref{eq_41}) or (\ref{eq_4-2}),
and the radius-mass relation (\ref{eq_31}) contain the same quantity $n$,
which means that they are related to each other.
One should stress  that the constants $\alpha$, $\beta$, and $\omega$ for planets, stars, galaxies, and clusters of galaxies differ
from each other.

For the star-like and planet-like objects  with  $n=3$,
Eq. (\ref{eq_6}) leads to $m=2/3$ and, correspondingly,
\begin{eqnarray}
R=\alpha M^{2/3},
\label{eq_311}
\end{eqnarray}
which is in the perfect agreement with the observational data
for the main sequence stars  \cite{Vasil:2012,Khal:2004,Cher:2013}.
For  main sequence stars, $\alpha=R_{\odot}/M_{\odot}^{2/3}$,
where $R_{\odot}$ and $M_{\odot}$ are the radius  and the mass of the Sun \cite{nash0}. 
From Eq. (\ref{eq_41}) we get
\begin{eqnarray}
\Im\sim M^{4/3}.
\label{eq_311Im}
\end{eqnarray}

For the galaxy-like or cluster of galaxy-like object with $n=2$,
we obtain $m=1/2$ from Eq. (\ref{eq_6})
or
\begin{eqnarray}
R=\alpha M^{1/2}
\label{eq_3112}
\end{eqnarray}
from Eq. (\ref{eq_31})
and
\begin{eqnarray}
\Im\sim M^{3/2}
\label{eq_8Im}
\end{eqnarray}
from Eq. (\ref{eq_41}).
Note that the derived $m=1/2$ is within  the observational range of data $m=\frac{2}{5} - \frac{2}{3}$ \cite{Karachentsev}.
For the galaxy-like objects, the definition of the constant $\alpha$ from the observational data is given in Ref. \cite{nash,nash1}.

%



\section{Neutron star and observable Universe}
\label{sec:Theory2}

Using Eq. (\ref{eq_1}), one can also obtain important
formulas for the exotic neutron star and the observable Universe, which are related, respectively,
to two important cosmological
Eddington and Chandrasekhar points on
 the cosmic analog of the Chew-Frautschi plot  \cite{Murad2,Murad3,Murad4}.
Equating $S$ in Eq. (\ref{eq_1}) to the Kerr maximal spin
\begin{eqnarray}
S=\hbar\left(\frac{M}{m_{Pl}}\right)^{2}=\hbar\left(\frac{M}{m_p}\right)^{2}I^{-1}=\frac{GM^2}{c},
\label{eq_K0}
\end{eqnarray}
of the rotating black hole ($I=\frac{\hbar c}{Gm_p^2}=1.69\times 10^{38}$ is the dimensionless combination of fundamental constants, $c$ is the speed of light in vacuum, and
$m_{Pl}=\left(\frac{\hbar c}{G}\right)^{1/2}=m_pI^{1/2}=1.3\times 10^{19}m_p=2.18\times 10^{-8}$ kg
is the Planck mass), one can derive
the Chandrasekhar mass
\begin{eqnarray}
M_{C}=m_pI^{3/2}
\label{eq_K1}
\end{eqnarray}
for $n=3$
and the Eddington mass
\begin{eqnarray}
M_{E}=m_pI^{2}
\label{eq_K22}
\end{eqnarray}
for $n=2$.
The resulting masses
$M_{C}=2.20\times 10^{57}m_p=3.46\times 10^{30}$ kg and
$M_{E}=2.87\times 10^{76}m_p=4.80\times 10^{49}$ kg
correspond  to
the Chandrasekhar  limiting mass of a degenerate neutron star
and the  Eddington  limiting mass  of the observable Universe, respectively \cite{Murad2,Murad3}.
As seen, the neutron mass $M_{C}$ is close to the  mass of the Sun ($M_{\odot}=1.99\times 10^{30}$ kg).

 Taking as the average masses of stars and galaxies in the Universe, respectively, the masses of the Sun
 and our Galaxy ($M_G=3.38\times 10^{41}$ kg) and using Eq. (\ref{eq_K22}),
we   roughly estimate the average number of galaxies
$$N_{\rm galaxy}\approx\frac{M_E}{M_G}\approx 10^8$$
and average number of stars
$$N_{\rm star}\approx\frac{M_E}{M_{\odot}}\approx 10^{19}$$
in the observable Universe.

The substitutions of (\ref{eq_K1}) and (\ref{eq_K22}) into  (\ref{eq_1}) or
(\ref{eq_K0}) lead to the limiting spins
\begin{eqnarray}
S_{C}=\hbar I^{2}
\label{eq_K3}
\end{eqnarray}
for the neutron star
and
\begin{eqnarray}
S_{E}=\hbar I^{3}
\label{eq_K4}
\end{eqnarray}
for the observable Universe.
Equation (\ref{eq_K4}) predicts the rotation of the entire astronomical Universe as a whole with the
spin 
$S_{E}=2.87\times 10^{76}\hbar=5.12\times 10^{80}$ J$\cdot$s  \cite{Murad2,Murad3}.
For the comparison,  
$S_{C}=2.20\times 10^{57}\hbar=3.03\times 10^{42}$ J$\cdot$s.
Employing the theoretical radius-mass relation
\begin{eqnarray}
R=\frac{GM}{c^2}
\label{eq_K5}
\end{eqnarray}
of  the rotating black hole and
Eqs. (\ref{eq_K1})  and  (\ref{eq_K22}),
one can derive \cite{Murad4} the
radius of neutron star
\begin{eqnarray}
R_{C}=r_pI^{1/2}
\label{eq_K6}
\end{eqnarray}
and
the radius  of the observable Universe
\begin{eqnarray}
R_{E}=r_pI,
\label{eq_K7}
\end{eqnarray}
where the factor  $r_p=\frac{\hbar}{m_pc}=2.10\times 10^{-16}$ m is the proton radius.
The calculated radii  (\ref{eq_K6}) and (\ref{eq_K7}) are the following:
$R_{C}=1.30\times 10^{19}r_p=2.74\times 10^{3}$ m
and
$R_{E}=1.69\times 10^{38}r_p=3.56\times 10^{22}$ m.
For comparison, the radius of the Sun is $R_{\odot}=6.99\times 10^{8}$ m.
The equality \cite{Murad3}
\begin{eqnarray}
\frac{S_E}{R_E^3}=\frac{\hbar}{r_p^3}
\label{eq_K67}
\end{eqnarray}
follows from Eqs. (\ref{eq_K4}) and (\ref{eq_K7}),
and implies that the spin densities of the proton and the Universe are the same within the factor 2.

Employing Eqs. (\ref{eq_K6}) and (\ref{eq_K7}), we obtain
 the following rotational frequencies
for the neutron star and the observable Universe ($\omega_p=\frac{c}{r_p}=1.43\times 10^{24} {\rm s}^{-1}$):
\begin{eqnarray}
\Omega_{C}&=&\frac{c}{R_{Ch}}=\frac{c}{r_p}I^{-1/2}=7.68\times 10^{-20}\frac{c}{r_p}=1.10\times 10^{5} {\rm s}^{-1},
\label{eq_K80}
\\
\Omega_{E}&=&\frac{c}{R_{E}}=\frac{c}{r_p}I^{-1}=5.90\times 10^{-39}\frac{c}{r_p}=8.43\times 10^{-15} {\rm s}^{-1}.
\label{eq_K8}
\end{eqnarray}
Note that
$$\frac{\Omega_{C}}{\Omega_{E}}=\frac{R_E}{R_C}=\frac{M_E}{M_C}=\left(\frac{S_E}{S_C}\right)^{1/2}=I^{1/2}=1.3\times 10^{19}.$$
The corresponding moments of inertia $\Im_{C}=2.76\times 10^{37}$ J$\cdot$s$^2$ and $\Im_{E}=6.08\times 10^{94}$ J$\cdot$s$^2$ are calculated as follows
\begin{eqnarray}
\Im_{C}=\frac{S_{C}}{\Omega_{C}}&=&\frac{\hbar r_p}{c}I^{5/2},\label{eq_K90}\\
\Im_{E}=\frac{S_{E}}{\Omega_{E}}&=&\frac{\hbar r_p}{c}I^{4},
\label{eq_K9}
\end{eqnarray}
where   $\frac{\hbar r_p}{c}=7.39\times 10^{-59}$ J$\cdot$s$^2$. Note that
$$\frac{\Im_{E}}{\Im_{C}}=\left(\frac{\Omega_{C}}{\Omega_{E}}\right)^{3}=\left(\frac{R_E}{R_C}\right)^{3}=\left(\frac{M_E}{M_C}\right)^{3}=\left(\frac{S_E}{S_C}\right)^{3/2}=I^{3/2}=2.2\times 10^{57}.$$
From Eqs. (\ref{eq_2}), (\ref{eq_K0}), and (\ref{eq_K5})  we obtain
$$\Im_{C,E}=\frac{S_{C,E}^2R_{C,E}}{\omega_{C,E} GM_{C,E}^2}=\frac{M_{C,E}R_{C,E}^2}{\omega_{C,E}}.$$
From comparing these equations with Eqs. (\ref{eq_K90}) and (\ref{eq_K9}), it follows that the dimensionless structural factors
for the neutron star and the observable Universe are equal to the unity, $\omega_{C}=\omega_{E} =1$, and $E_k=\beta=0$.
Surprisingly, the moments of inertia (\ref{eq_K90}) and (\ref{eq_K9})
are larger than corresponding rigid-body moments of inertia.

Counting the average age of the Universe on the order of 14 billion years  and using Eq. (\ref{eq_K8}),
we derive a numerical value of the angular velocity of rotation of the Universe
\begin{eqnarray}
\Omega_E=4\times10^3 \frac{1}{{\rm age}-{\rm of}-{\rm Universe}}.
\label{eq_K10}
\end{eqnarray}
In other words, the total rotation time of the  observable  Universe is approximately 10$^{7}$ years,
which is about of 10$^{3}$ times less than the average age of the Universe.
In Ref. \cite{Murad3},  the estimated frequency estimate is about 6 orders of magnitude less than the value given by Eq. (\ref{eq_K10})
due to the difference in the moments of inertia used in the calculations.


\section{Darwin instability effect in binary systems}
\label{sec:Theory3}

Now let us apply the Regge-theory  to the astronomical compact binary systems.
When the mass ratio in the binary compact star is extreme enough for the Darwin instability \cite{Darwin:1879},
a merger of the binary components starts that triggers  the outburst in a red novae \cite{Tylenda:2011}.
The Darwin instability happens when the spin
 of the system is more than one third of the orbital angular momentum.
This instability plays a role
once the mass ratio became  small enough that the companion star can
no longer keep the primary star synchronously rotating via the tidal
interaction.
Angular momentum transferred from the binary orbit to the
  intrinsic spin changes the orbit and leads
to a runaway.
For most of the primary massive stars,
this occurs at the mass ratio $q=M_2/M_1< 0.1$ \cite{Rasio:1995}.

The total angular momentum  ${\bf J}$  of the  binary system is the sum of
orbital angular momentum ${\bf L}$ and the spins ${\bf S}_k$ ($k=$1, 2) of the individual components:
\begin{eqnarray}
{\bf J} = {\bf L} + {\bf S}_1 + {\bf S}_2.
\label{eq_9b}
\end{eqnarray}
The $J$ and $S_k$ are expressed using the Regge-like laws \cite{Murad1,Murad2,Murad3}:
\begin{eqnarray}
J=\hbar\left(\frac{M}{m_p}\right)^{1+1/n}
\label{eq_10b}
\end{eqnarray}
and
\begin{eqnarray}
S_k=\hbar\left(\frac{M_k}{m_p}\right)^{1+1/n},
\label{eq_11b}
\end{eqnarray}
where  $M_k$ ($k=$1,2)    and $M=M_1+M_2$ are
masses of   the binary components  and the total mass of the system, respectively.
Then, the maximum (the antiparallel orbital and spin  angular momenta) and minimum (the  parallel orbital and spin  angular momenta)
 orbital angular momenta are
\begin{eqnarray}
L_{\rm max}=J+S_1+S_2
\label{eq_12b}
\end{eqnarray}
and
\begin{eqnarray}
L_{\rm min}=J-S_1-S_2,
\label{eq_13b}
\end{eqnarray}
respectively.
%
Using  Eqs. (\ref{eq_9b})--(\ref{eq_13b}), we derive
\begin{eqnarray}
\frac{S_1+S_2}{L_{\rm min}}=\frac{1+q^{1+1/n}}{(1+q)^{1+1/n}-q^{1+1/n}-1}
\label{eq_60b}
\end{eqnarray}
and
\begin{eqnarray}
\frac{S_1+S_2}{L_{\rm max}}=\frac{1+q^{1+1/n}}{(1+q)^{1+1/n}+q^{1+1/n}+1}.
\label{eq_14b}
\end{eqnarray}
At $q=1$ (the symmetric binary system) and $n>1$, we have
$$\frac{S_1+S_2}{L_{\rm min}}=\frac{1}{2^{1/n}-1}>1$$ and
$$\frac{S_1+S_2}{L_{\rm max}}=\frac{1}{2^{1/n}+1}>\frac{1}{3}.$$
For the symmetric binary star or binary planet ($n=3$), or   binary galaxy ($n=2$) with $q=1$,
 $(S_1+S_2)/L_{\rm max}\approx$ 0.44 and 0.41, respectively.
At $q\to 0$, we have
$$\frac{S_1+S_2}{L_{\rm min}}\to\infty$$ and
$$\frac{S_1+S_2}{L_{\rm max}}\to\frac{1}{2}.$$
As follows from these two expressions,  for very asymmetric binary system,
the ratios $(S_1+S_2)/L_{\rm max,min}$ are almost independent of the value of $n$.
According to Ref.  \cite{Rasio:1995}, the Darwin instability can occur  when the binary mass ratio is very small  [$q=M_2/M_1 < 0.1$] or the
mass asymmetry  is very large.
The ratios $(S_1+S_2)/L_{\rm max}$ and $(S_1+S_2)/L_{\rm min}$
 continuously increase   with decreasing $q$   from 1 to 0.
Because their absolute values are larger than 1/3,  all possible binary stars or binary planets, or binary galaxies, independently of their mass ratios $q$,
should have the Darwin instability ($S_1 + S_2 \ge \frac{1}{3}L$) \cite{Darwin:1879} and, correspondingly, should merge.
However,  the observations do not support this conclusion
which probably  means that there is no   Darwin instability effect in such binary systems and, correspondingly, the mechanism of merging
has other origin.
Thus, it is necessary to search for another mechanism that triggers
the merging of the contact binary components.


As follows,   in the cases of    antiparallel spins with $L_{12}=J+S_1-S_2$   and
$L_{21}=J-S_1+S_2$, the ratios $|S_2-S_1|/L_{12}$ and $|S_1-S_2|/L_{21}$
are larger than $\frac{1}{3}$ for the asymmetric binaries with $q\le 1/3$ \cite{nash2}.

\section{Summary}
\label{sec:Summary}
Within   the model \cite{Murad1,Murad2,Murad3} based on the concept of  Regge trajectories,
 we have derived the new analytical formulas  for the moment of inertia [Eq. (\ref{eq_41})], the rotation frequency [Eq. (\ref{eq_51})], and the radius
 [the relation (\ref{eq_6}) between $m$ and $n$ or Eqs. (\ref{eq_31}), (\ref{eq_311}), (\ref{eq_3112})] of astronomical   objects
 (stars, planets, galaxies, and clusters  of galaxies).
The moment of inertia   (\ref{eq_41})  depend  on the total  mass $M$,   dimension $n$ of the object, the dimensionless structural factor $\omega$, the
classical and quantum fundamental constants $G$, $\hbar$, $m_p$,
and the constants  $\alpha$ and $\beta$ extracted from the observed data.

The formulas for the rotation frequencies [Eqs. (\ref{eq_K80}) and (\ref{eq_K8})]
and moments of inertia  [Eqs. (\ref{eq_K90}) and (\ref{eq_K9})] of the neutron
star and the observed Universe are derived.
The estimate of the speed of rotation of the Universe [Eq. (\ref{eq_K10})]
is about (5--7) orders of magnitude larger than the that from previous estimates.
The  rotation time of the  observable  Universe is about 10$^{7}$ years.
The estimated  average numbers of galaxies and  stars in the observable Universe are
$\sim 10^8$ and $\sim 10^{19}$, respectively.

Employing the  Regge-like laws,
we have also shown that all possible binary stars (binary planets) or binary galaxies
satisfy the Darwin instability condition ($S_1 + S_2 \ge \frac{1}{3}L$) which contradicts to the observations.
This conclusion is not sensitive to the parameters of model.
Therefore, one should search for other mechanism that triggers
the merger of the contact binary components.

\vspace{3cm}
This work was partially supported by  Russian Foundation for Basic Research (grant 20-02-00176, Moscow)   and
DFG (grant Le439/16, Bonn). V.V.S. acknowledges the Alexander von Humboldt-Stiftung (Bonn).



\begin{thebibliography}{0}

\bibitem{Regge} T. Regge, Nuovo Cimento {\bf 14} (1959) 951.

\bibitem{Regge2} A. Bottino, A. M. Longoni, and  T. Regge, Il Nuovo Cimento {\bf 23} (1962) 954.

\bibitem{CFplot} G. F. Chew and S. C. Frautschi, Phys. Rev. Lett. {\bf 7} (1961) 394.

\bibitem{Murad1} R. M. Muradian,  Astrofiz. {\bf 11} (1975) 237 [in Russian];
Astrofiz. {\bf 13} (1977) 63 [in Russian]; Astrofiz. {\bf 14} (1978) 439 [in Russian].

\bibitem{Murad2} R. M. Muradian,  Astrophys. Space Sci. {\bf 69} (1980) 339.

\bibitem{Murad3} R. M. Muradian,  Phys. Part. Nucl. {\bf 28} (1997) 471.

\bibitem{Ambart} V. A. Ambartsumian,  {\it  La Structure et I'Evolution de I'Universe}, Solvay conference  (Brussels),
 {\bf 11} (1958) 241.

\bibitem{Ambart1} V. A. Ambartsumian,  Astrophys. J. {\bf 66} (1961) 536.

\bibitem{Ambart2}  V. A. Ambartsumian, {\it  The Structure and  Evolution of Galaxies},  Solvay conference  (Brussels),   {\bf 13}  (1965) 1.

\bibitem{Darwin:1879} G. H. Darwin,  Proc. R. Soc. {\bf 29} (1879) 168.

\bibitem{Vasil:2012}
 B. V.   Vasiliev,  Univ. J. Phys. Applic. {\bf 2} (2014) 257; {\bf 2} (2014) 284; {\bf 2} (2014) 328;
 J. Mod. Phys. {\bf 9} (2018) 1906; {\bf 9} (2018) 2101.

\bibitem{Khal:2004}
 K. F.  Khaliullin, {\it  Ph. D. Thesis}, Sternberg Astronomical Institute, Moscow (2004).



\bibitem{Cher:2013} A. M.  Cherepashchuk, {\it  Close binary stars}, Fizmatlit, Moscow (2013), volumes I and  II
 [in Russian].



\bibitem{nash0} V. V. Sargsyan, H. Lenske, G. G. Adamian, and N.V.   Antonenko,
Int. J. Mod. Phys. E {\bf 27} (2018)  1850063; {\bf 27} (2018) 1850093; {\bf 28} (2019) 1950044.


\bibitem{Karachentsev} I.D. Karachentsev, {\it Binary galaxies},
Nauka, Moscow (1987) [in Russian].


\bibitem{nash} V. V. Sargsyan, H. Lenske, G. G. Adamian, and N. V.   Antonenko,
Int. J. Mod. Phys. E {\bf 28} (2019) 1950031.

\bibitem{nash1} V. V. Sargsyan, H. Lenske, G. G. Adamian, and N. V.   Antonenko,
  Russ. J. Nucl. Phys. {\bf 83} (2020) 61.

\bibitem{Murad4} R. M. Muradian, S. Carneiro, and  R. Marques, arXiv: astro-ph/9907129v1 (1999).

\bibitem{Tylenda:2011}
R. Tylenda  {\it et al.},    Astron. Astrophys. {\bf 528} (2011) A114.

\bibitem{Rasio:1995} F. A. Rasio,   ApJL {\bf 444} (1995)  L41.

\bibitem{nash2}  G. G. Adamian,  N. V.   Antonenko,  H. Lenske, and V. V. Sargsyan,
Commun. Theor. Phys.  {\bf 71} (2019) 1335.

\end{thebibliography}
\end{document}